\documentclass[12pt,a4paper,twoside]{article}

\usepackage{epsfig}
\usepackage{baltlat}
\begin{document}
\ \ 
\vspace{0.5mm}

\setcounter{page}{1}
\vspace{8mm}

\titlehead{Baltic Astronomy, vol.11, ***--***, 2002.}

\titleb{THE CONSEQUENCES OF ASSUMING $m$=0 FOR \\ GLOBAL MODEL-FITTING}

\begin{authorl}
\authorb{Travis S.~Metcalfe}{1}
\end{authorl}

\begin{addressl}
\addressb{1}{Theoretical Astrophysics Center, Aarhus University, DENMARK}
\end{addressl}

\submitb{Received August 5, 2002}

\begin{abstract}
A recent re-analysis of Whole Earth Telescope observations of GD~358
obtained in 1990 suggests that asteroseismology of additional DBV white
dwarfs can lead to independent constraints on the important, but poorly
determined, $^{12}{\rm C}(\alpha,\gamma)^{16}{\rm O}$ nuclear reaction
rate. Data exist for several other DBV white dwarfs, but relatively few
modes are detected and there is often no multiplet structure to aid in
the identification of the spherical harmonic indices ($\ell$, $m$). I use 
a new grid of one million DBV models covering a broad range of masses,
temperatures, and surface helium layer masses to investigate the
consequences of assuming $m$=0 for global model-fitting. I find that when
the spherical degree is known and the rotation period is of order 1 day,
the model-fitting procedure applied to modes with unknown $m$-values will
still correctly identify the {\it families} of possible solutions, and 
has a high probability of identifying the same globally optimal solution 
found when the $m$-value is known.
\end{abstract}


\resthead{Assuming $m$=0 for global model-fitting}{T.S.~Metcalfe}

\sectionb{1}{INTRODUCTION}

More than a decade after the outstanding successes with PG~1159 and GD~358
(Winget et al.~1991, 1994), the Whole Earth Telescope (WET; Nather et
al.~1990) continues to look for variable white dwarfs with rich pulsation
spectra. Unfortunately we have learned that these stars are truly
exceptional---either in the number and variety of their intrinsic
variations, or in our ability (and luck) to detect so many modes. A more
typical WET run tends to yield a smaller number of modes, not necessarily
with consecutive radial overtone numbers ($k$), and with few (if any)
triplets or quintuplets to simplify the identification of the spherical
degree ($\ell$) and azimuthal order ($m$) of each mode. This observational
difficulty limits our ability to find meaningful theoretical model-fits
for other white dwarfs. Consequently, what we have learned from more
recent runs has been limited when compared to the shining successes of the
past.

\sectionb{2}{DBV MODEL GRID}

Recent work using the 1990 data for GD~358 has suggested that reliable
measurements of the central ratio of carbon to oxygen can be derived from
asteroseismological data, which can subsequently yield precise constraints
on the rate of the important $^{12}{\rm C}(\alpha,\gamma)^{16}{\rm O}$
nuclear reaction (Metcalfe et al.~2000, 2001, 2002). The analysis of
additional DBV white dwarfs using the same model-fitting method can, in
principle, provide independent constraints on the reaction rate under
varying conditions of density and temperature. However, until recently
none of the other DBV white dwarfs had a sufficient number of observed
periods, and the mode identifications were not secure. This meant that, at
best, only the mass, temperature, and surface helium layer mass could be
reliably deduced from the limited data.

With a larger number of free parameters a genetic algorithm based approach
is preferable, but the prospect of fitting many DBVs with only 3
parameters makes the calculation of a full grid of models more efficient
in the long run. Using a 32-processor beowulf system at Aarhus University,
the $\ell$=1, $m$=0 pulsation periods between 100 and 1000 seconds were
calculated for one million carbon-core DBV models with masses between 0.45
and 0.95 $M_{\odot}$ (0.005 $M_{\odot}$ resolution), temperatures between
20,000 and 30,000 K (100 K resolution), and $\log(M_{\rm He}/M_*)$ between
$-$2.0 and $-$7.3 (0.05 dex resolution). The grid required nearly 4
GHz-CPU-months to complete, but it was finished on the cluster after only
a few days. This model grid now makes it possible to map the entire
3-dimensional parameter-space for a given set of observed periods in about
30 seconds\footnote[1]{A web interface to the model-fitting program that 
uses this grid is available online at {\tt 
http://whitedwarf.org/research/fitwd/}}. Such a map for GD~358 in 1990 is 
shown in Figure 1.

\begin{figure}[t]
\centerline{\psfig{figure=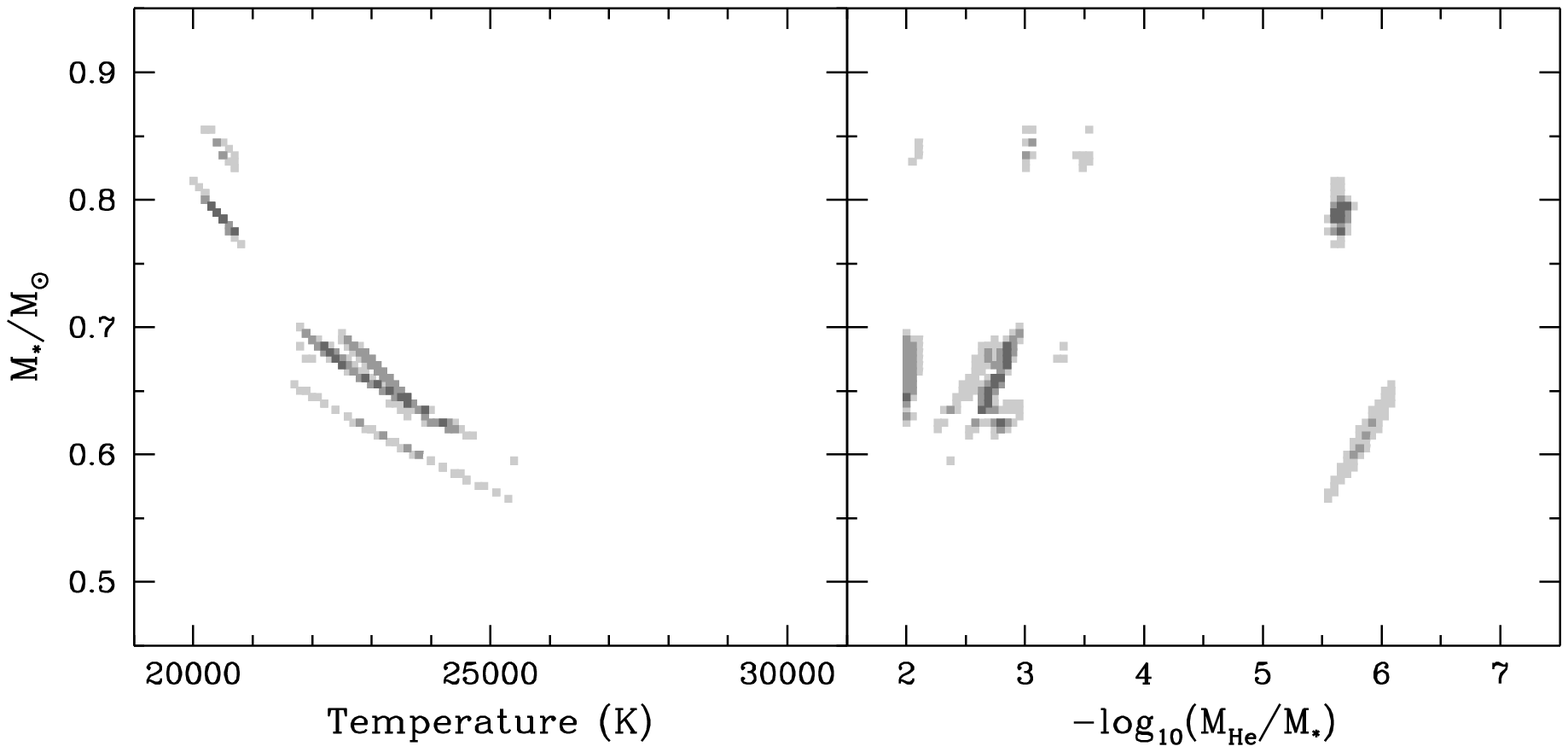,width=12truecm,angle=0,clip=}}
\captionb{1}{Front and side views of the model-space for the pulsation
periods observed in GD~358 showing the families of good matches to the
observations. Each point in the left panel corresponds one-to-one with a
point in the right panel. The shade of each point is an indication of how
well the model matches the observations: root-mean-square period
differences are less than 3 seconds (lightest), 2.75 seconds, and 2.5
seconds. High mass models can be ruled out based on the luminosity
of GD~358 from its observed parallax.}
\vskip2mm
\end{figure}

\sectionb{3}{MONTE CARLO SIMULATIONS}

Single-site observations of the faint DBV white dwarf CBS~114 revealed a
total of 7 independent pulsation modes, and the period spacing implies
that they all have spherical degree $\ell$=1 (Handler, Metcalfe, \& Wood
2002). None of the modes exhibited multiplet structure, so the
identification of the $m$-value was not possible. In 1990, triplets were
observed in 9 of the 11 identified modes for GD~358, so we can use this
data to determine empirically the consequences of making a wrong
assumption about the $m$-values.

One hundred random data sets were created, drawing the period for each $k$
from the observed $m$=($-$1,0,+1) triplets at random. Since no triplets
were observed for $k$=12 and $k$=18, these periods were included unaltered
into each simulated data set. The model parameters that yielded the
closest match to each set of simulated periods were determined by
comparison with the entire grid of models described in section 2, and the
root-mean-square period residuals were calculated.

\begin{figure}
\centerline{\psfig{figure=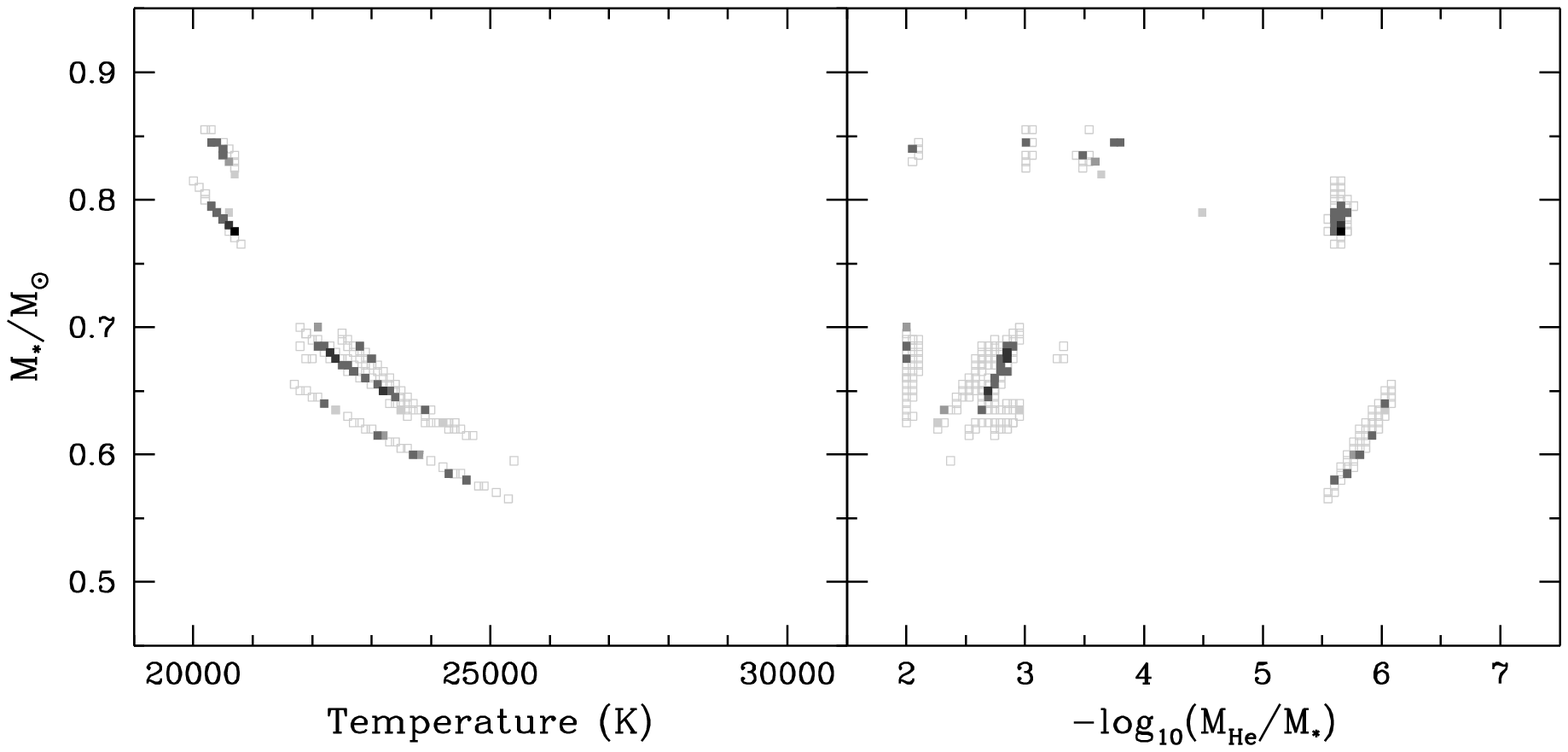,width=12truecm,angle=0,clip=}}
\captionb{2}{The distribution of globally optimal model parameters for 
the 100 simulated data sets created using random $m$-values from the observed
modes in GD 358. The families of models from the $m$=0 case in Figure 2 are
shown in outline for comparison. Again the shade indicates root-mean-square
period differences less than 3 seconds (lightest), 2.75, 2.5, 1.75, and 
1.5 seconds (darkest).}
\vskip2mm
\end{figure}

\sectionb{4}{RESULTS \& DISCUSSION}

The globally optimal models for each simulated data set are shown in
Figure 2, where the darkness of the point indicates the quality of the
fit. The families of models from Figure 1 are outlined for reference. In
nearly every case, the data sets created from random $m$-values lead to
optimal models that fall within the same families of models found when all
of the modes are $m$=0. The basic picture of the model-space that we
derive from modes with unknown $m$-value is very similar to what we see when
they are all known to be $m$=0. The {\it particular} set of model
parameters that turns out to be optimal may shift slightly, or may even
fall into a different family of models when the $m$-values are unknown.
Furthermore, the optimal models fall into the various families in rough
proportion to the average quality of the fits in each region. In more than
80 percent of the simulations, the optimal model fell near the two darkest
areas in Figure 1. Thus, we can think of the inclusion of modes with
non-zero values of $m$ as a perturbation on the shape of the model-space:
the general terrain is determined by the spacing of the modes regardless
of their $m$-value, and small local features on that terrain are determined
by the correct identification of $m$.

The rotation period of GD~358 at the surface is 0.9 days (Winget et 
al.~1994), so these results are only applicable to white dwarfs with
comparably slow rotation. But it is encouraging that without a secure
identification of $m$, we can still be confident that our overall picture
of the model-space is sound. We can map out the possibilities, and to the
degree that the optimal model is unique we can maximize the probability of
finding an optimal solution very near to what it would be if we were
certain that all of the modes were $m$=0.
\vskip7mm

ACKNOWLEDGEMENTS.\ I would like to thank Mike Montgomery for suggesting
this empirical test, and Gerald Handler for providing the motivation, 
in the form of CBS~114. This work was supported by the Danish National 
Research Foundation through its establishment of the Theoretical
Astrophysics Center.
\goodbreak

\References

\ref{Handler, G., Metcalfe, T.S., \& Wood, M.A. 2002, MNRAS, in press}

\ref{Metcalfe, T.S. et al. 2000, ApJ, 545, 974}

\ref{Metcalfe, T.S. et al. 2001, ApJ, 557, 1021}

\ref{Metcalfe, T.S. et al. 2002, ApJ, 573, 803}

\ref{Nather, R.E. et al. 1990, ApJ, 361, 309}

\ref{Winget, D.E. et al. 1991, ApJ, 378, 326}

\ref{Winget, D.E. et al. 1994, ApJ, 430, 839}

\end{document}